\documentstyle[preprint,aps,prd,epsf]{revtex}

\def\overleftrightarrow#1{\vbox{\ialign{##\crcr
    $\leftrightarrow$\crcr\noalign{\kern-1pt\nointerlineskip}
    $\hfil\displaystyle{#1}\hfil$\crcr}}}
\def\dbw{\overleftrightarrow\partial}

\begin{document} 
\draft 
\tighten

\preprint{\vbox{\hbox{SUSX-TH/97-017}}}

\title{Electrically Charged Sphalerons.}

\author{ P.M. Saffin\thanks{E-mail: p.m.saffin@sussex.ac.uk}\& 
         E.J. Copeland\thanks{E=mail: E.J.Copeland@sussex.ac.uk}} 
\address{Centre for Theoretical Physics, University of
Sussex, \\Brighton BN1 9QH, United Kingdom}

\date{\today}
\maketitle

%@twocolumnfalse\endcsname

\begin{abstract} 
\noindent We investigate the possibility that the Higgs sector 
of the Weinberg-Salam model admits the existence of electrically
charged, sphaleron states. Evidence is provided through an asymptotic and 
numerical perturbative analysis about the uncharged sphaleron. By 
introducing a toy model in two dimensions we demonstrate 
that such electrically charged, unstable states can exist. 
Crucially, they can have a comparable mass to their uncharged counterparts 
and so may also play a role in electroweak 
baryogenesis, by opening up new channels for baryon number violating processes.
\end{abstract}
%\vskip2pc]

\section{Introduction}
\noindent There has for some time been great interest in the mechanism
which generated the baryon anti-baryon asymmetry in the early Universe.
An important observation was made by t'Hooft \cite{thooft}, when he 
noted that the Adler-Bell-Jackiw anomalies
\cite{abj} in the standard model meant that baryon number was not conserved
quantum mechanically. This opened up the possibility 
that baryogenesis could occur naturally at the
electroweak scale. Moreover, the fact that sphalerons formed in the 
early Universe possessed the ability to wash
out any effect of primordial baryogenesis \cite{russian_chaps}, focussed 
attention on the electroweak transition as the source for the 
observed baryon--asymmetry.  
The field configuration now known as the sphaleron first appeared 
in the work of Daschen et al \cite{dhn}, but its potential role 
in baryon number violating transitions
was only later clarified by Manton \cite{manton}. 
It is a saddle point solution between
field configurations of baryon number zero and one, 
whilst itself has a baryon number of one half. 

The energy of the sphaleron is of the order of 10Tev such that at 
comparable temperatures sphaleron configurations may be thermally excited, 
leading to baryon number violation.
Interest in a charged state with fractional baryon number arises
from the possibility that a new class of decay channels
could emerge which would lead to baryon number
violation if they were mediated by such a charged sphaleron, 
namely electrically charged initial states.
One reason why we might expect such a solution on physical grounds comes
from an interpretation of the SU(2) sphaleron. This state may be considered
as a pair of oppositely charged magnetic monopoles in a special unstable
equilibrium, where the attractive magnetic forces are balanced by repulsive
topological ones. Vachaspati \cite{tanmay} has indicated 
the existence of dyon
states in the standard model and it seems a natural extension to consider
an unstable equilibrium of two dyons rather than monopoles.

If we are to investigate the existence of a charged 
sphaleron then we cannot work
in the symmetry breaking scheme SU(2)$\rightarrow$1, 
where the sphaleron
was first found, as there is no electric
charge symmetry in the vacuum of this model. 
We must therefore consider the full 
SU(2)$\times$U(1)$\rightarrow$U(1) of the standard model. 
This case is complicated
by the presence of currents which reduce the 
sphaleron symmetry
from being spherically to axially symmetric, leading to a 
magnetic dipole moment
\cite{markymark}.
As far as we are aware, such states have not yet been observed in numerical
simulations of the non abelian phase transition . Two reasons for this are 
that most dynamical simulations take
a vanishing Weinberg angle \cite{turok} or, in the
case of static calculations, force the temporal components of the gauge 
fields to vanish \cite{kkb}. 
Both of which effectively forbid 
the existence of such states.

\section{Toy Model}
To investigate the charge on the sphaleron we start by introducing a
toy model which has an electromagnetic symmetry generator in the
vacuum. 
The main motivation for introducing such a model is that the equations 
obtained are ordinary rather than partial differential equations, allowing
a simpler numerical treatment.
In three spatial dimensions the symmetry breaking pattern
\mbox{SU(2)$\rightarrow$U(1)} leads to the topologically stable 
t'Hooft-Polyakov monopoles. In two dimensions however this admits
static, unstable states which are the analog of the sphalerons in the
standard model. Our lagrangian for a gauged SU(2) theory is,
%%%%%%%%%%%%%%%%%%%%%%%%%%%%%%
\begin{eqnarray}
{\cal L}&=&\frac{1}{2}(D_{\mu} \Phi^a)(D^{\mu} \Phi^a)
           -\frac{1}{4}W^{a \mu \nu}W^a_{\mu \nu}
           -{\cal V}(\Phi^a\Phi^a),
\end{eqnarray}
%%%%%%%%%%%%%%%%%%%%%%%%%%%%%%
where
%%%%%%%%%%%%%%%%%%%%%%%%%%%%%%
\begin{eqnarray}
\Phi^a&=&(\Phi^1,\Phi^2,\Phi^3) \\
D_{\mu} \Phi^a&=&\left(\partial_{\mu}\Phi^a + g \epsilon^{abc}W^{b}_{\mu}  \Phi^c \right) \\
W^a_{\mu \nu}&=&\partial_\mu W^a_\nu-\partial_\nu W^a_\mu
                +g \epsilon^{abc} W^{b}_{\mu} W^{c}_{\nu} 
\end{eqnarray}
%%%%%%%%%%%%%%%%%%%%%%%%%%%%%%
with $g$ as the gauge coupling constant.
The two dimensional aspect of the theory means that the 
single winding, uncharged sphaleron is found 
using a Nielsen-Olesen string
ansatz where the relevant gauge field is \mbox{$A_\mu=W^3_\mu$} with 
\mbox{$W^1_\mu=W^2_\mu=0$} and only the upper two components of the triplet 
higgs field are used: $\phi^{1(2)}=\Phi^{1(2)}$, $\Phi^3=0$. It is convenient 
to introduce the unit vector $n^a$ and to re-express the Higgs and 
gauge fields,  
%%%%%%%%%%%%%%%%%%%%%%%%%%%%%%
\begin{eqnarray}
n^i&=&(x/r,y/r), \\
\phi^a&=& \eta h(r) n^a/g \\
A^{i}&=&f(r) \epsilon^{ij}n^j/(gr).
\end{eqnarray}
%%%%%%%%%%%%%%%%%%%%%%%%%%%%%%
This ansatz may then be extended to include charged states by allowing 
for some of the temporal components of the gauge fields to be non-zero,
%%%%%%%%%%%%%%%%%%%%%%%%%%%%%%
\begin{eqnarray}
W^{a0}&=&J(r)n^a/g\\
W^{30}&=&0.
\end{eqnarray}
%%%%%%%%%%%%%%%%%%%%%%%%%%%%%%
With this, the profile functions $h(r)$, $f(r)$, $J(r)$ are found to satisfy,
%%%%%%%%%%%%%%%%%%%%%%%%%%%%%%
\begin{eqnarray}
h''+h'/r-h/r^2&=&2hf(1+f/2)/r^2+2h\frac{\partial {\cal V}}{\partial \phi^2} \\
f''-f'/r&=&(1+f)(\eta^2h^2-J^2) \\
J''+J'/r-J/r^2&=&2Jf(1+f/2)/r^2.
\end{eqnarray}
%%%%%%%%%%%%%%%%%%%%%%%%%%%%%%
These equations are solved using a relaxation technique, where the 
profiles have the following boundary conditions. At the origin, 
$h(r\rightarrow 0)\rightarrow a r$,
$f(r\rightarrow 0)\rightarrow b r^2$,
$J(r\rightarrow 0)\rightarrow c r$, whereas asymptotically we find,
$h(r\rightarrow \infty)\rightarrow 1$,
$f(r\rightarrow \infty)\rightarrow -1$,
$J(r\rightarrow \infty)\rightarrow A+B\ln(r)$.
The presence of a logarithmic term should come as no surprise since Laplace's 
equation for electrostatics in two dimensions has a logarithmic solution.
The electromagnetic field tensor is defined by 
\mbox{${\cal F}_{\mu \nu}=\phi^a W^a_{\mu \nu}/|\phi|$}, and the 
electric field is
then $E^i={\cal F}^{0i}=\partial_i J(r)/g$. To find the charge enclosed in 
a circle of radius $r$, the electric field is integrated over the circumference 
to 
give $Q(r)=\int d\sigma^i E^i$. The asymptotic form of $J(r)$ then shows that,
$Q(r\rightarrow \infty) \rightarrow 2 \pi B/g=Bg/\alpha$, 
where $\alpha=g^2/(2 \pi)$ is the 
fine structure constant in two dimensions. In particular the total 
charge enclosed is finite. 
Typical profiles are shown below in Fig1.

An important property of sphalerons is their mass. 
Since they are thermally activated, their creation probability 
is suppressed by a 
Boltzmann factor. If the charged states are excessively massive
then this exponential suppression would render their formation rate negligible.
Using the canonical definition of the stress-energy-momentum tensor,
$T^{\mu \nu}=-2/(\sqrt(-g)) \delta S/ \delta g_{\mu \nu}$, with S being the action,
we define the mass as $m=\int d^2 x T^{00}$. The variation of mass as a 
function of charge
is shown in Fig2. For small charges the mass rises slowly as the 
charge is increased. This behavior is reminiscent of the energy-charge 
relation for the dyon in the BPS limit \cite{bog}, E$_{dyon} \propto \sqrt(1+q^2) $
 where
$q \propto Q$, the charge. From the presence of the fine structure 
constant we see that
for a sphaleron to have a charge comparable to the elementary charge then the mass
cost is small and so the Boltzmann factor for these charged sphaleron states is
not significantly greater than the uncharged states, making them physically 
interesting and worth investigating in the full three dimensional picture. 

\section{Asymptotic Analysis in the Standard Model.}
We now turn our attention to the full symmetry breaking pattern of
the standard model. Although it is too difficult to obtain complete analytic 
solutions for the charged state sphaleron, we intend to demonstrate the 
existence of the solution through a combination of numerical and analytical 
means. We start by determining the asymptotics
of a static, compact field configuration. This technique 
was used by Vachaspati
\cite{tanmay} when looking at the construction of dyons in the standard model,
but will be included here for completeness.
The lagrangian of the higgs sector of the standard model is written as,
%%%%%%%%%%%%%%%%%%%%%%%%%%%%%%
\begin{eqnarray}
{\cal L}&=&(D_{\mu} \Phi)^{\dagger}(D^{\mu} \Phi)
           -\frac{1}{4}W^{a \mu \nu}W^a_{\mu \nu}
           -\frac{1}{4}B^{\mu \nu}B_{\mu \nu}
           -{\cal V}(\Phi^{\dagger}\Phi).
\end{eqnarray}
%%%%%%%%%%%%%%%%%%%%%%%%%%%%%%
where,
%%%%%%%%%%%%%%%%%%%%%%%%%%%%%%
\begin{eqnarray}
\label{dmuphi}
D_{\mu} \Phi&=&\left(\partial_{\mu} - g W^{a}_{\mu} t^a -{i\over 2} 
g' B_{\mu} \right) \Phi\\
\left[t^a,t^b\right]&=&-\epsilon^{a b c} t^c\\
\left[D_{\mu},D_{\nu}\right]\Phi&=&-\left(gW_{\mu \nu}+i{g'\over 2} 
B_{\mu \nu}\right)\Phi\\
D_{\mu} W^{a \sigma \mu}&=&\partial_{\mu} W^{a \sigma \mu}
                          +g\epsilon^{a b c} W^{b}_{\mu} W^{c \sigma \mu},
\end{eqnarray}
%%%%%%%%%%%%%%%%%%%%%%%%%%%%%%
with $B_\mu$ an abelian gauge field with coupling constant $g'$. The corresponding 
field equations are,
%%%%%%%%%%%%%%%%%%%%%%%%%%%%%%
\begin{eqnarray}
D^\mu D_\mu \Phi &=&-\frac{\partial {\cal V}}{\partial \Phi^{\dagger}}\\
\partial_{\mu} B^{\sigma \mu}&=&i/2 g' \left[\Phi^{\dagger} D^{\sigma} \Phi
                                         -(D^{\sigma} \Phi)^{\dagger}\Phi \right]\\
D_{\mu} W^{a \sigma \mu}&=& g \left[\Phi^{\dagger} t^a D^{\sigma} \Phi
                                         -(D^{\sigma} \Phi)^{\dagger} t^a \Phi \right].
\end{eqnarray}
%%%%%%%%%%%%%%%%%%%%%%%%%%%%%%
Finite energy considerations imply that 
asymptotically $D_{\mu}\Phi=0$. From Eq.~(\ref{dmuphi}) and using the appendix we find 
%%%%%%%%%%%%%%%%%%%%%%%%%%%%%%
\begin{eqnarray}
\label{split}
-g W^a_{\mu}
+g' B_{\mu}n^a&=&-in^a\left(\Phi^{\dagger}\dbw_\nu\Phi\right) 
   +\epsilon^{a b c} n^b \partial n^c\\
n^a&=&2i\Phi^{\dagger} t^a \Phi/\Phi^{\dagger}\Phi,
\end{eqnarray}
%%%%%%%%%%%%%%%%%%%%%%%%%%%%%%
where $\underline{n}$
is a unit vector. It proves useful to split Eq.~(\ref{split}) 
into parts parallel and perpendicular to $\underline{n}$, with 
the general solution,
%%%%%%%%%%%%%%%%%%%%%%%%%%%%%%
\begin{eqnarray}
-g W_{\mu}^a&=&\epsilon^{abc}n^b\partial_\mu n^c
               -i \cos^2(\alpha) n^a \left(\Phi^{\dagger}\dbw_\mu\Phi\right)
               +n^a a_\mu\\
g'B_{\mu}&=&-i\sin^{2}(\alpha)\left(\Phi^{\dagger}\dbw_\mu\Phi\right)-a_{\mu},
\end{eqnarray}
%%%%%%%%%%%%%%%%%%%%%%%%%%%%%%
where $a_\mu$ is arbitrary and represents the gauge freedom 
of the statement $D_{\mu}\Phi=0$ \cite{manton_monopole}, and $\alpha$ is 
determined by the dynamics of the 
particular field configuration we are investigating. 
As we are looking to charge the sphaleron 
then we are particularly interested in $a_0$.
We now suppose that there is a static solution, denoted by barred fields, 
representing the usual uncharged sphaleron from which we may construct a new, charged
state.
Now choose the lorentz vector \mbox{$a_{\mu}=\gamma \delta^{0}_{\mu}$}
such that,
%%%%%%%%%%%%%%%%%%%%%%%%%%%%%%
\begin{eqnarray}
\Phi&=&\overline{\Phi}\\
W^a_{\mu}&=&\overline{W}^a_{\mu}
           -\frac{1}{g} \gamma n^a \delta^0_{\mu}\\
B_{\mu}&=&\overline{B}_{\mu}-\frac{1}{g'}\gamma\delta^0_{\mu}.
\end{eqnarray}
%%%%%%%%%%%%%%%%%%%%%%%%%%%%%%
By construction $D_{\mu} \Phi= \overline{D}_\mu \overline{\Phi}=0$ so
$\Phi$ satisfies the usual scalar field equation. The 
space-space components of the field strengths remain unchanged and the 
space-time components are found to be,
%%%%%%%%%%%%%%%%%%%%%%%%%%%%%%
\begin{eqnarray}
B_{0i}&=&\frac{1}{g'}\partial_i \gamma\\
W^a_{0i}&=&-\left(\partial_i W^a_0+g\epsilon^{abc}\overline{W}^b_i W^c_0\right)\\
        &=&\frac{1}{g} \overline{D}_i (\gamma n^a).
\end{eqnarray}
%%%%%%%%%%%%%%%%%%%%%%%%%%%%%%
As we are in the vacuum, $D_\mu \Phi=0$, then clearly $D_i (n^a)=0$ 
which means 
that the field equations for $B_{\mu}$ and $W^a_{\mu}$ 
each lead to the relations,
%%%%%%%%%%%%%%%%%%%%%%%%%%%%%%
\begin{eqnarray}
\partial_i \partial^i \gamma&=0\\
\partial_0 \partial^i \gamma&=0.
\end{eqnarray}
%%%%%%%%%%%%%%%%%%%%%%%%%%%%%%
A separation of variables then shows 
$\gamma(\underline{x},t)=\gamma(\underline{x})$ 
and $\nabla^2 \gamma=0$. With $\gamma$ satisfying 
Laplaces equation we therefore know that
asymptotically the time component of the gauge fields take the form,
%%%%%%%%%%%%%%%%%%%%%%%%%%%%%%
\begin{eqnarray}
W^a_0&=&\frac{1}{g} n^a 
        \sum^{l=\infty}_{l=0} \sum^{m=l}_{m=-l} 
        C_{l,m} P^m_l(\cos(\theta))/r^{l+1}\\
B_0&=&\frac{1}{g'} 
      \sum^{l=\infty}_{l=0} \sum^{m=l}_{m=-l} 
        C_{l,m} P^m_l(\cos(\theta))/r^{l+1}.
\end{eqnarray}
%%%%%%%%%%%%%%%%%%%%%%%%%%%%%%
Now, the electromagnetic and Z potentials are given by,
%%%%%%%%%%%%%%%%%%%%%%%%%%%%%%
\begin{eqnarray}
A_{\mu}&=& \sin(\theta_w) n^a W^a_{\mu}+\cos(\theta_w) B_{\mu}\\
\label{zmu}
Z_{\mu}&=& \cos(\theta_w) n^a W^a_{\mu}-\sin(\theta_w) B_{\mu},
\end{eqnarray}
%%%%%%%%%%%%%%%%%%%%%%%%%%%%%%
where $\theta_w$ is the weinberg angle. 
From eq.~(\ref{zmu}) we see that the $Z_{\mu}$ potential vanishes at infinity, 
as is fitting for a massive
field. However the electromagnetic potential, $A_0$, being a solution to 
Laplace's equation is able to support all multipole moments, including the 
zeroth monopole term giving rise
to electric charge -- the charge that we are interested in.
As for the baryonic charge of the state, this is found to be \cite{manton},
%%%%%%%%%%%%%%%%%%%%%%%%%%%%%%
\begin{eqnarray}
Q_B&=&=\frac{g^2}{32 \pi^2} \int d^3 x K^0\\
\nonumber K^\mu&=&\epsilon^{\mu\nu\rho\sigma}
                  \left( W^a_{\nu\rho}W^a_{\sigma}
                        -\frac{2}{3}g\epsilon_{abc}W^a_\nu W^b_\rho W^c_\sigma
                  \right).
\end{eqnarray}
%%%%%%%%%%%%%%%%%%%%%%%%%%%%%%
When we look at the zero component of $K$ to find the 
sphaleron's baryonic charge,
only the space components of the gauge field are used. These 
have remained unaltered
from the uncharged sphaleron, consequently the charged and 
uncharged sphaleron have the same baryonic charge.

\section{Numerical Analysis of the Charged Sphaleron.}
In the previous section we analyzed the 
asymptotic behaviour of the fields. 
To investigate the core of the charged sphaleron we first look at
the general axisymmetric field configuration to see what 
form the temporal components of the gauge fields
can take. Then we shall use the SU(2) 
sphaleron as a background in which we numerically
investigate whether a charged sphaleron state exists for small $\theta_w$.
The question of spatial symmetry in gauge theories has a subtlety in as much as a 
gauge transformation is not a physical transformation and 
we are concerned only
with physical symmetries. A configuration is axisymmetric if a rotation about
an axis is equivalent to a gauge transformation \cite{kkb},
%%%%%%%%%%%%%%%%%%%%%%%%%%%%%%
\begin{eqnarray}
\left({\cal G}+{\cal O}\right)\Xi&=&0,
\end{eqnarray}
%%%%%%%%%%%%%%%%%%%%%%%%%%%%%%
where $\Xi$ represents a generic field and ${\cal G}$, ${\cal O}$ are the 
generators of gauge transformations and rotations respectively.
To help with the generation of a suitable ansatz we introduce 
the set of orthonormal vectors
%%%%%%%%%%%%%%%%%%%%%%%%%%%%%%
\begin{eqnarray}
\nonumber \underline{u}_x&=&\left(x/\rho,y/\rho,0\right)\\
\nonumber \underline{u}_y&=&\left(-y/\rho,x/\rho,0\right)\\
\underline{u}_z&=&\left(0,0,1\right)\\
\nonumber R^2&=&x^2+y^2+z^2\\
\nonumber \rho^2&=&x^2+y^2.
\end{eqnarray}
%%%%%%%%%%%%%%%%%%%%%%%%%%%%%%
It may then be shown \cite{kkb} that the following field configuration is annihilated
by a sum of rotation and gauge transformation generators.
%%%%%%%%%%%%%%%%%%%%%%%%%%%%%%
\begin{eqnarray}
\nonumber W^a_i(\underline{x})&=&u^i_j u^a_k w^k_j(\rho,z)\\
B_i(\underline{x})&=&u^a_i d_a(\rho,z)\\
\nonumber \Phi(\underline{x})&=&h_i(\rho,z)u^a_i t^a \Psi\\
\nonumber \Psi&=&\left(0,\eta/\sqrt 2\right),
\end{eqnarray}
%%%%%%%%%%%%%%%%%%%%%%%%%%%%%%
a result that guarantees the axisymmetric nature of the configurations. 
This system of fields may be extended to include possible zero
components of the gauge fields thus,
%%%%%%%%%%%%%%%%%%%%%%%%%%%%%%
\begin{eqnarray}
\nonumber W^a_0(\underline{x})&=&u^a_i F^i(\rho,z)\\
B_0(\underline{x})&=&J(\rho,z).
\end{eqnarray}
%%%%%%%%%%%%%%%%%%%%%%%%%%%%%%
It is possible to restrict the situation further to solutions which
are invariant under various discrete symmetries. We impose the condition 
that the combined transformation of charge conjugation, parity reflection 
and the SU(2) transformation -1 leaves our sphaleron unchanged \cite{kkb}.
This boils down to,
%%%%%%%%%%%%%%%%%%%%%%%%%%%%%%
\begin{eqnarray}
\nonumber w^x_x=w^y_y=w^z_z=w^x_z=w^z_x=0\\
F_y=0\\
\nonumber d_x=d_z=0\\
\nonumber h_y=0.
\end{eqnarray}
%%%%%%%%%%%%%%%%%%%%%%%%%%%%%%
The numerical investigation is based on a perturbative expansion for small
$g'/g$. In this case the $w^a_\alpha$ and $h_x$, $h_z$ are found from
the $g'/g=0$ sphaleron, namely,
%%%%%%%%%%%%%%%%%%%%%%%%%%%%%%
\begin{eqnarray}
W^{ai}=2 f(R) \epsilon^{aij}\frac{x^j}{R^2}\\
\nonumber \Phi=2 h(R) \frac{x^i}{R} t^i \Psi,
\end{eqnarray}
%%%%%%%%%%%%%%%%%%%%%%%%%%%%%%
where $h(R)$ and $f(R)$ satisfy,
%%%%%%%%%%%%%%%%%%%%%%%%%%%%%%
\begin{eqnarray}
\frac{\partial^2 f}{\partial R^2}&=&
       2\frac{f}{R^2}(1+gf)(1+2gf)+\frac{1}{4}g\eta^2 h^2 (1+gf)\\
R^2 \nabla^2 h&=&2h(1+gf)^2
                 +R^2 h \frac{\partial {\cal V}}{\partial (\Phi^{\dagger} \Phi)}
\end{eqnarray}
%%%%%%%%%%%%%%%%%%%%%%%%%%%%%%
The only non-vanishing part of $d_\alpha$, $d_y$, has 
been investigated previously
\cite{markymark}, and leads to a magnetic dipole moment. 
The form of $B_i$ in the small $g'/g$ limit is found to be
\begin{eqnarray}
B_i&=&-g'p(R)\epsilon_{ij3}x_j\\
\frac{\partial^2 p}{\partial R^2}+\frac{4}{R}\frac{\partial p}{\partial R}
    &=&-\frac{1}{2}\eta^2 h^2 \left(1+gf\right)
\end{eqnarray}
The profiles of the functions involved in the small $g'/g$ uncharged 
sphaleron are shown below in Fig3.

The
terms $J$, $F_x$, $F_z$ determine the sphaleron charge.
We have solved for them numerically
in the background of the sphaleron with vanishing Weinberg angle.
The definition of the electromagnetic tensor ${\cal F}_{\mu \nu}$ 
and currents used here
are the same as in \cite{markymark},
\begin{eqnarray}
{\cal F}_{\mu \nu}&=&\sin(\theta_W)n^a W^a_{\mu \nu}+\cos(\theta_W)B_{\mu \nu}\\
\partial^\nu {\cal F}_{\mu \nu}&=&j_\mu.
\end{eqnarray}
The resulting electric charge distribution, $j_0$, is displayed in Fig4 where we see
clear evidence 
that the electric charge density is localized at the centre of the system.

The issue of electric charge quantization is a tricky one. The charged sphaleron, 
having two
displaced regions of magnetic charge density will develop
an angular momentum, due to the Poynting vector. 
If this angular momentum is then quantized we shall
find ourselves with sphalerons that occur with discrete values of electric
charge. Such a calculation has successfully been performed in the case of 
a charged topological string in an abelian higgs model
\cite{zee}, where exact analytic expressions are obtained. 
Unfortunately, the precise nature of the quantization in our case is obscured by the
complexity of the field equations and the fact that we do not have exact analytic 
results to use. The principle though still holds, evaluating the angular momentum for 
different charged states should lead to a determination of the fundamental electric 
charge.  

\section{Conclusion.}
The aim of this paper has been to demonstrate that there may exist 
unstable charged states in the standard model. These could contribute to the process
of electroweak baryogenesis. By analogy with a sphaleron
in two dimensions we showed that such states do exist. Following Vachaspati
\cite{tanmay} we saw that asymptotically the fields can support a charged
distribution and a perturbative numerical treatment also indicated that such 
charged states exist.
Finally, it was shown how the axial symmetry of the sphaleron, which causes the
computational problems leads to a natural way to quantize the electric charge.

When considering how such configurations would contribute to baryogenesis 
we need to know the mass of such 
states. It was shown that the 
mass-charge relation of the 2D sphaleron was similar to that of the BPS dyon
in that the mass increases slowly for small, physical values of electric charge. 
If this relation is generic, which may be expected as the uncharged sphaleron is
so massive, then the charged sphalerons in the standard model will be excited
at temperatures similar to the uncharged states. Such states would open up a new 
channel for baryon violating processes in the standard model through their interactions 
with charged particles. This may prove to be a fruitful avenue of investigation.

%%%%%%%%%%%%%%%%%%%%%%%%%%%%%%%%%%%%%%%%%%%%%%%%%%%%%%%%%%%%%%
\acknowledgments
We are grateful to M. Hindmarsh, T. Barreiro, and J. Grant for useful conversations. 
EJC and PS are 
grateful to PPARC for financial support. Partial support was obtained from the 
European Commission under the Human Capital and Mobility program, contract no. 
CHRX-CT94-0423. 

%%%%%%%%%%%%%%%%%%%%%%%%%%%%%%%%%%%%%%%%%%%%%%%%%%%%%%%%%%%%%%
 
\listoffigures

\noindent Figure 1. A plot of the higgs field and SU(2) vector field
                    profiles for the two dimensional sphaleron.

\noindent Figure 2. A plot showing the variation of mass with total charge
                    for the two dimensional sphaleron.

\noindent Figure 3. A plot of the higgs field and SU(2) vector field
                    profiles for the zero Weinberg angle sphaleron.

\noindent Figure 4. A plot showing the density of electric charge
                     associated with the sphaleron.
\newline

%%%%%%%%%%%%%%%%%%%%%%%%%%%%%%%%%%%%%%%%%%%%%%%%%%%%%%%%%%%%%%%%%%%%%%%%%%%%%%
\begin{figure}[!htb]
\centering
\mbox{\epsfxsize=300pt\epsfysize=300pt\epsffile{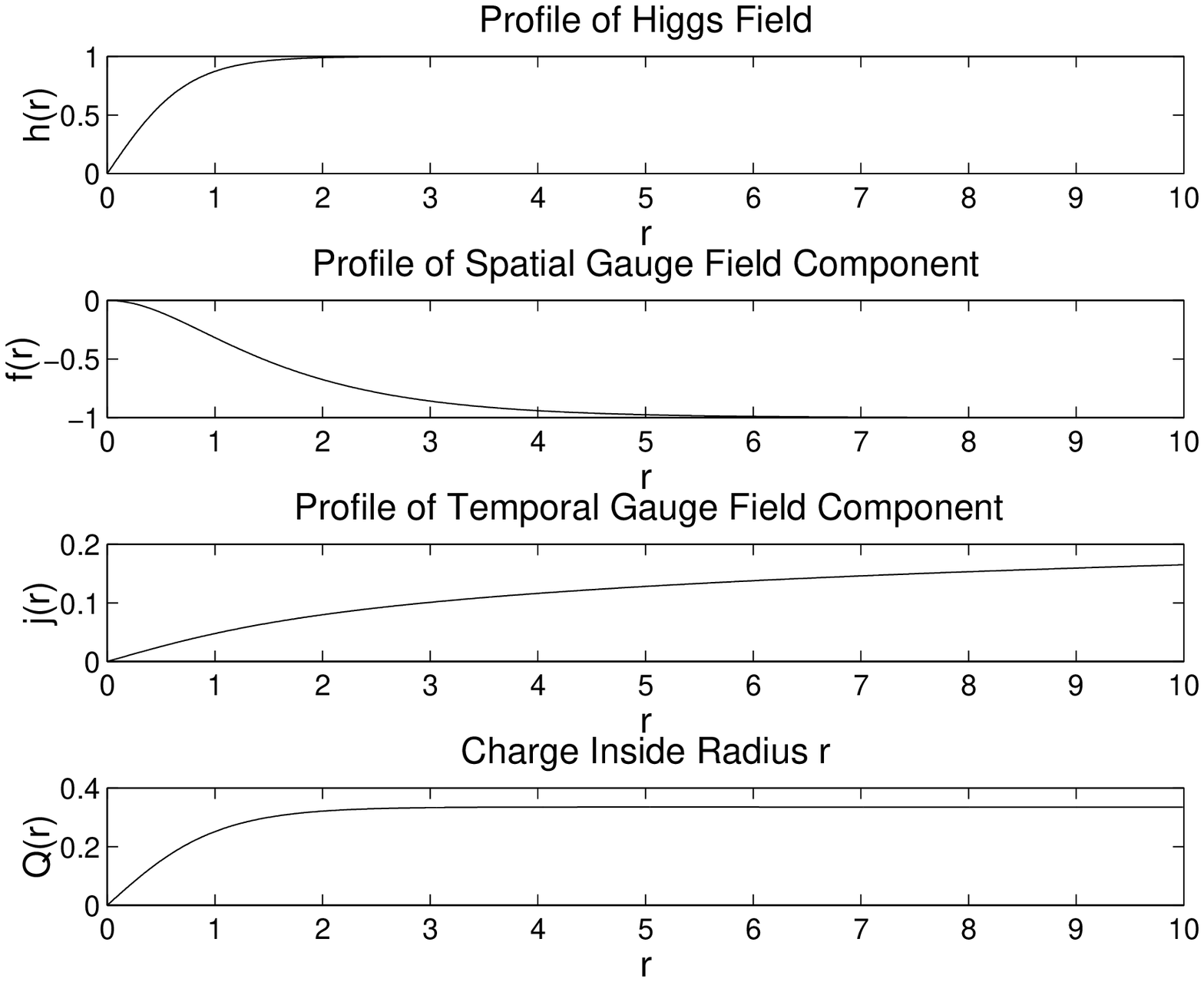}}
\end{figure}
%%%%%%%%%%%%%%%%%%%%%%%%%%%%%%%%%%%%%%%%%%%%%%%%%%%%%%%%%%%%%%%%%%%%%%%%%%%%%%
\begin{figure}[!htb]
\centering
\mbox{\epsfxsize=300pt\epsfysize=300pt\epsffile{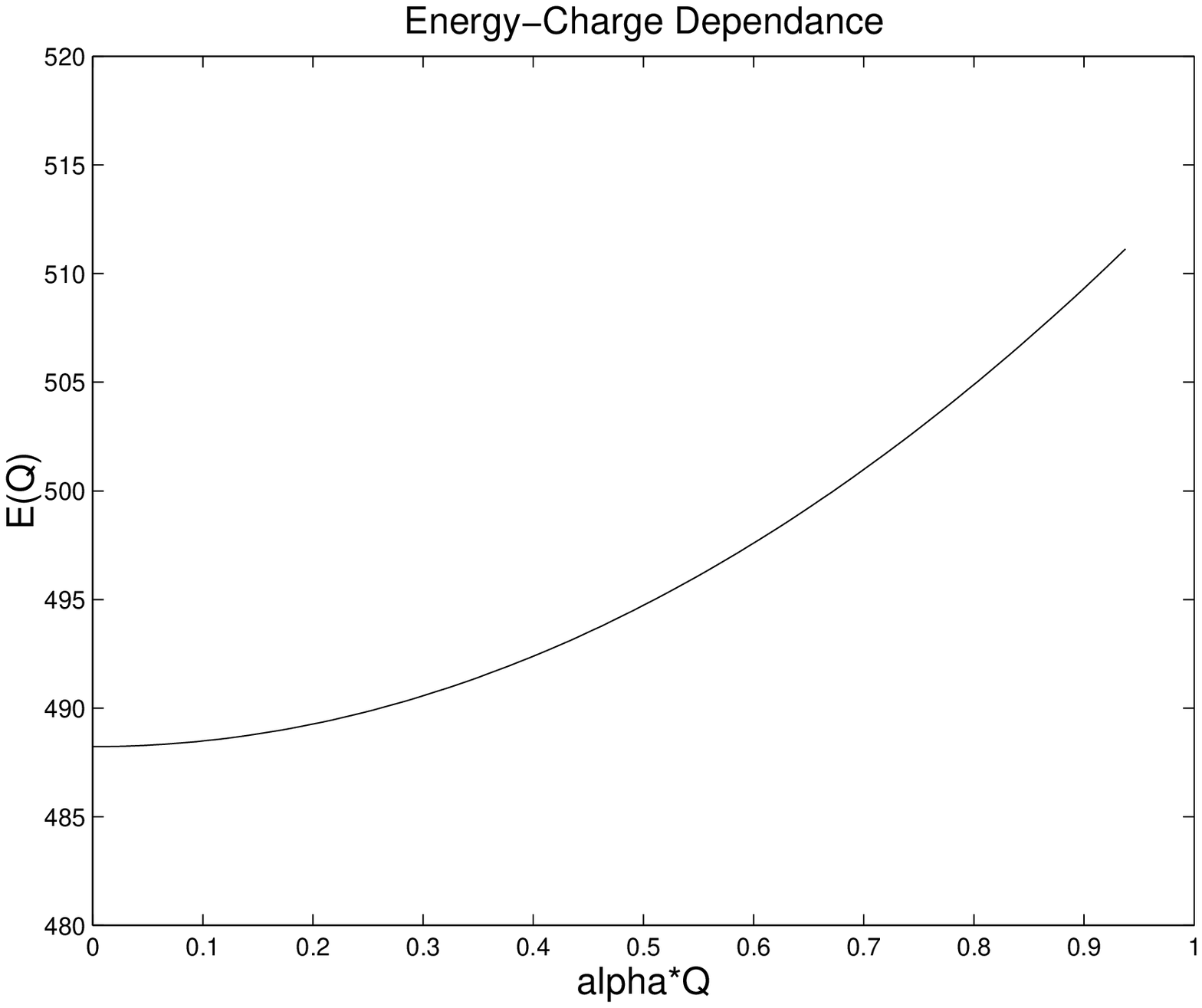}}
\end{figure}
%%%%%%%%%%%%%%%%%%%%%%%%%%%%%%%%%%%%%%%%%%%%%%%%%%%%%%%%%%%%%%%%%%%%%%%%%%%%%%
\begin{figure}[!htb]
\centering
\mbox{\epsfxsize=300pt\epsfysize=300pt\epsffile{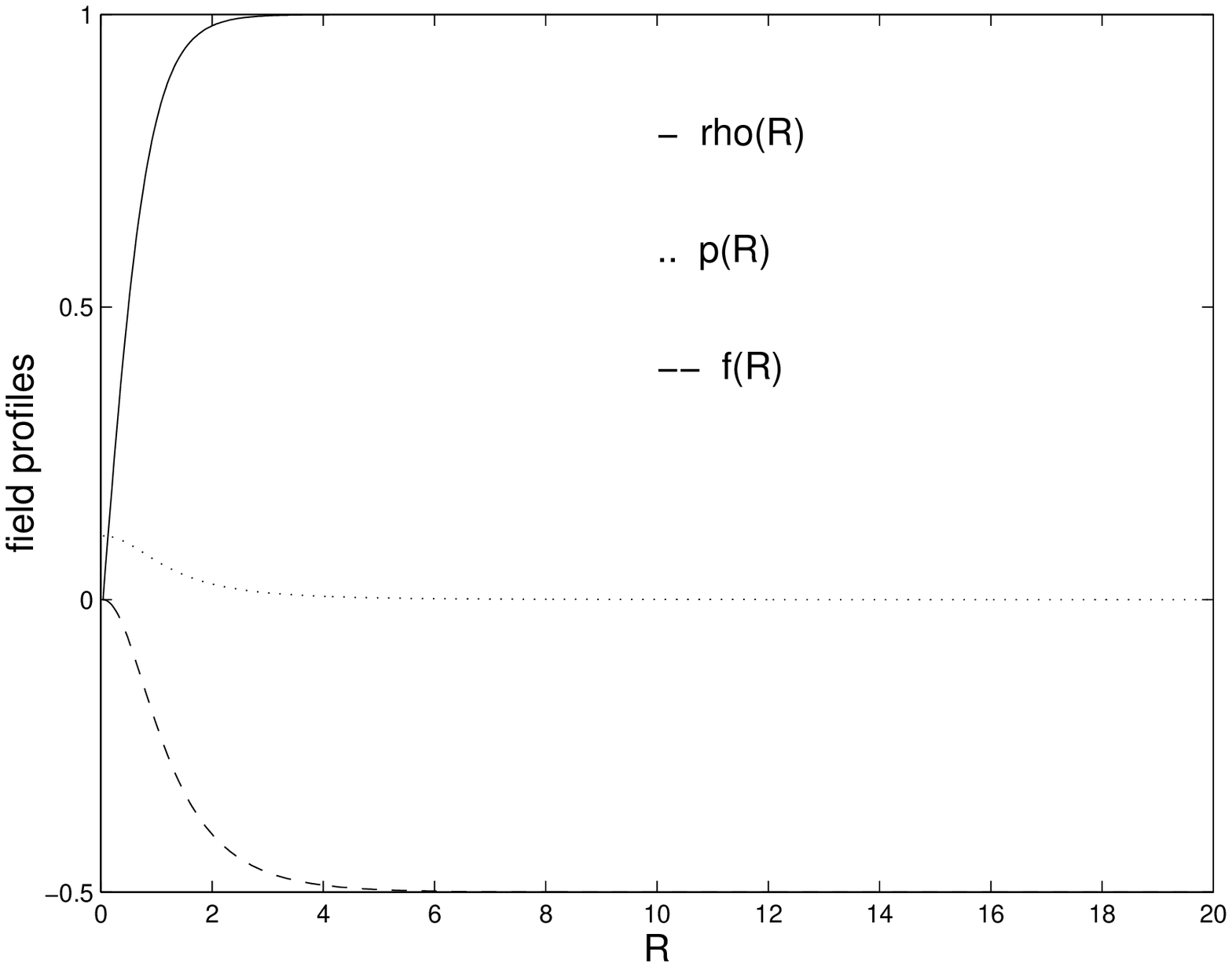}}
\end{figure}
%%%%%%%%%%%%%%%%%%%%%%%%%%%%%%%%%%%%%%%%%%%%%%%%%%%%%%%%%%%%%%%%%%%%%%%%%%%%%%
\begin{figure}[!htb]
\centering
\mbox{\epsfxsize=300pt\epsfysize=300pt\epsffile{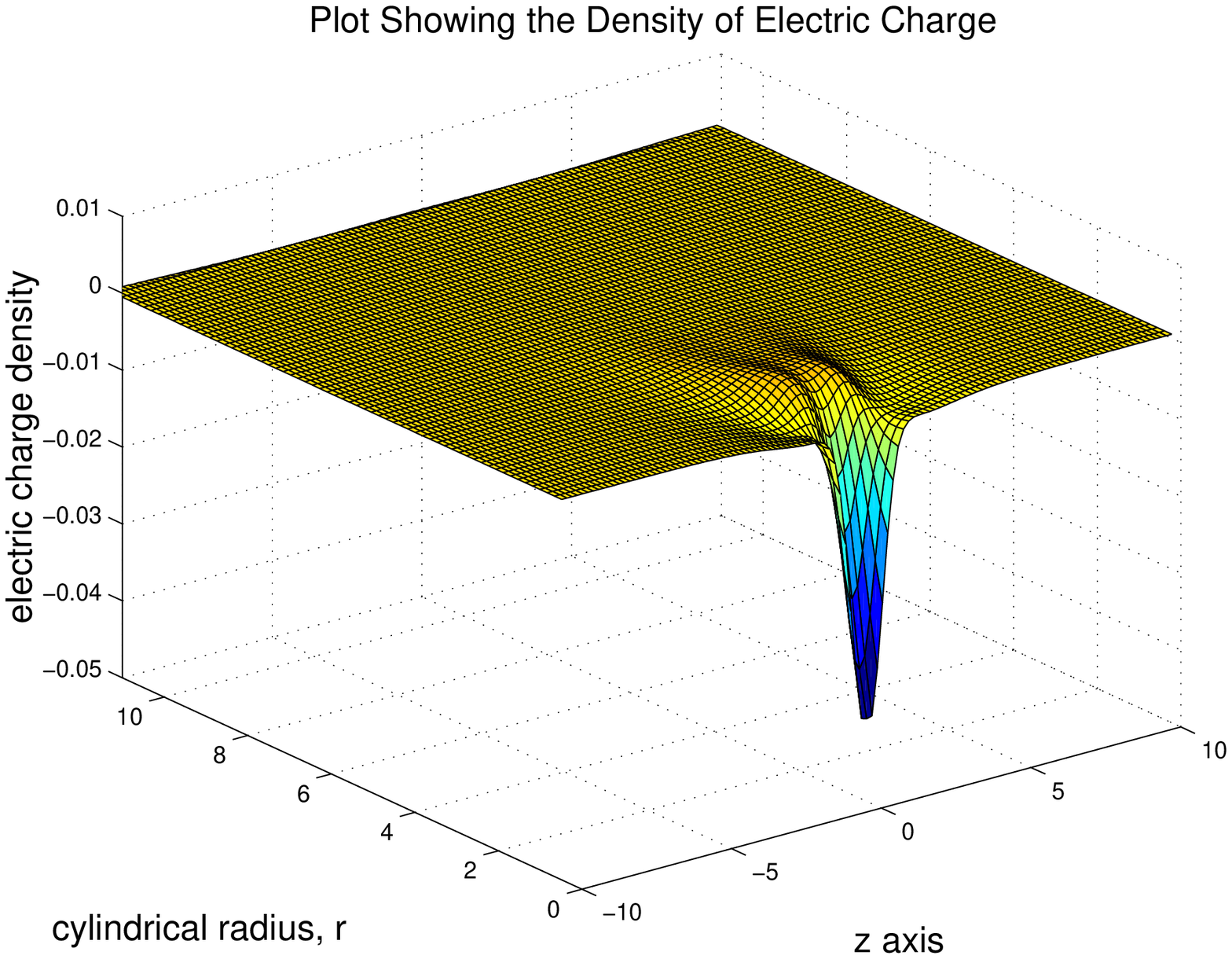}}
\end{figure}
%%%%%%%%%%%%%%%%%%%%%%%%%%%%%%%%%%%%%%%%%%%%%%%%%%%%%%%%%%%%%%%%%%%%%%%%%%%%%%
\appendix
\section*{}
Consider the set of matrices $\Gamma^{\mu}=(1,2it^a)$. 
These then satisfy
$Tr\left(\Gamma^{\mu} \Gamma^{\nu} \right)=2\delta^{\mu \nu}$. 
Using the fact that the $\Gamma^{\mu}$ form a linearly independent basis of 2x2 
complex matrices, we see that 
$$\Gamma^{\mu}_{a b} \Gamma^{\mu}_{\beta \alpha}=
      2\delta_{a \alpha}\delta_{b \beta},$$
where summation over repeated indices is implied.
Then by considering
\begin{itemize}
\item $\left(\Phi^{\dagger}\Gamma^{\mu}\Phi\right) 
       \left(\Phi^{\dagger}\Gamma^{\mu}\Phi\right)$
\item $\left(\Phi^{\dagger}\Gamma^{\mu}\Phi\right)\Gamma_{\mu}\Phi$
\item $\left(\Phi^{\dagger}\Gamma^{\mu}\Phi\right) 
       \left(\Phi^{\dagger}\Gamma^{\mu}\partial_{\nu}\Phi\right)$
\item $\left(\Phi^{\dagger}\Gamma^{\mu}\Phi\right) 
       \partial_{\nu}\left( \Phi^{\dagger}\Gamma^{\sigma}\Gamma^{\mu}\Phi\right)$
\end{itemize}
we find that
\begin{itemize}
\item $\left(\Phi^{\dagger}t^a\Phi\right) 
       \left(\Phi^{\dagger}t^a\Phi\right)
       =-1/4\left(\Phi^{\dagger}\Phi\right)^2$
\item $\left[1+4(\Phi^{\dagger} t^a \Phi )/\Phi^{\dagger}\Phi \right]\Phi
       =0$
\item $\left(\Phi^{\dagger} t^a \Phi\right) 
       \left(\Phi^{\dagger} t^a \partial_{\nu}\Phi\right)
       =-1/4 \left(\Phi^{\dagger}\Phi\right)
             \left(\Phi^{\dagger}\partial_{\nu}\Phi\right)$
\item $\left(\Phi^{\dagger}\Phi\right)\left(\Phi^{\dagger} t^a \dbw_\nu\Phi\right) 
       =\left(\Phi^{\dagger} t^a \Phi\right)\left(\Phi^{\dagger}\dbw_\nu\Phi\right)
    -2\epsilon^{a b c}\left(\Phi^{\dagger} t^b \Phi\right)
           \partial_{\nu}\left(\Phi^{\dagger} t^c \Phi\right),$
\end{itemize}
expressions which prove useful in the asymptotic analysis of the charged 
sphaleron solutions.

\end{document}